\documentclass[aps,pra,twocolumn,amsmath,showpacs,showkeys,superscriptaddress,longbibliography,floatfix]{revtex4-1}

\usepackage{amsmath}
\usepackage{amssymb}
\usepackage{physics}
\usepackage{epsfig}
\usepackage{graphicx}
\usepackage{graphics}
\usepackage{url}
\usepackage[english]{babel}
\usepackage{dcolumn}
\usepackage{color}
\usepackage[squaren]{SIunits}
\usepackage{hyperref}
\usepackage{cleveref}

\crefname{section}{Sec.}{Sections}
\crefname{equation}{Eq.}{Eqs.}
\Crefname{equation}{Equation}{Equations}
\crefname{figure}{Fig.}{Figs.}
\Crefname{figure}{Figure}{Figures}

\begin{document}

\newcommand{\mean}[1]{\bar{#1}}
\newcommand{\vect}[1]{\boldsymbol{#1}}
\newcommand{\erw}[1]{\langle {#1} \rangle}
\newcommand{\ska}[2]{\langle {#1}, {#2} \rangle}
\newcommand{\highi}[2]{#1^{(#2)}}
\newcommand{\te}[1]{\text{#1}}
\newcommand{\Epsilon}[0]{\mathcal{E}}

\title{Semiclassical modeling of coupled quantum dot-cavity systems: From polariton-like dynamics to Rabi oscillations}

\author{K.~J\"urgens}
\author{F.~Lengers}
\author{T.~Kuhn}
\author{D.~E.~Reiter}
\affiliation{Institut f\"ur Festk\"orpertheorie, Universit\"at M\"unster,
Wilhelm-Klemm-Str.~10, 48149 M\"unster, Germany}

\date{\today}

\begin{abstract}
Semiconductor quantum dots in photonic cavities are strongly coupled light-matter systems with prospective applications in optoelectronic devices and quantum information processing.
Here we present a theoretical study of the coupled exciton--light field dynamics of a planar quantum dot ensemble, treated as two-level systems, embedded in a photonic cavity modeled by Maxwell's equations.
When excited by coupling an external short laser pulse into the cavity, we find an exciton-polariton-like behavior for weak excitation and Rabi oscillations for strong excitation with a sharp transition between these regimes.
In the transition region we find highly non-linear dynamics involving high harmonics of the fundamental oscillation.
We perform a numerical study based on the Finite-Difference-Time-Domain method for the solution of Maxwell's equations coupled to Bloch equations for the quantum dots and also derive an analytical model to describe the coupled cavity-quantum dot system, which allows us to describe the light field dynamics in terms of a Newton-like dynamics in an effective anharmonic potential.
From the shape of this potential combined with the initial conditions the transition can be well understood.
The model is then extended to a broadened ensemble of quantum dots.
For weak excitation the polariton spectrum broadens and the lines slightly shift, however, the sharp transition to the Rabi oscillation regime is still present.
Furthermore, we find a second, lower threshold with additional lines in the spectra which can be traced back to Rabi oscillations driven by the polariton modes.
Our approach provides new insights in the dynamics of both quantum dot and light field in the photonic structure. 
\end{abstract}

\pacs{XXX} 


\maketitle

\section{Introduction}
In state-of-the-art semiconductor nanostructures the electronic and optical properties can be tailored by spatial confinement of the electronic and/or photonic degrees of freedom.
The ultimate electronic confinement is reached in semiconductor quantum dots (QDs) which, due to the three-dimensional confinement of the electronic states, have a discrete energy spectrum.
These QDs can be embedded in photonic structures like micro cavities \cite{reitzenstein2007ala,schneider2016qua,laucht2009dep,muller2015ult}, photonic crystal structures \cite{englund2005con,lodahl2004con}, nano lenses \cite{gschrey2015hig,fischbach2017sin} or plasmonic structures \cite{deinega2014int}.
By confining the light modes these structures change the local density of photon states leading to an increase of the light-matter coupling with the QDs \cite{takeda2011sel}. This is a crucial aspect, when considering QDs to be used as single or entangled photon sources \cite{heindel2017bri,senellart2017hig,huber2018sem}, but also for optoelectronic devices which are based on QD ensembles used, e.g., for lasing applications \cite{czerniuk2017pic}.

To model such structures it is required to account for both, the dynamics of the QD system and the photonic structure.
In the limit of strongly confined light, popular approaches are based on the Jaynes-Cummings model from cavity-quantum-electrodynamics \cite{brune1996qua,yoshie2004vac,khitrova2006vac,laussy2008str}, which typically accounts for one quantized light mode and a single two-level system, but not explicitly for the spatio-temporal light-field dynamics in the photonic structure.
For QD-systems the Jaynes-Cummings model is often extended by considering additionally the electron-phonon interaction \cite{hughes2011inf,moelbjerg2012res,kaer2013mic,cosacchi2019emi,reiter2019dis}.
Instead of looking at a single QD in a microcavity, here we consider a planar ensemble of QDs in a one-dimensional photonic cavity formed by a pair of Bragg mirrors \cite{reitzenstein2007ala,schneider2016qua}.
In order to study the dynamics of the system we combine the equations of motion for the QD exciton with a Finite-Difference-Time-Domain (FDTD) method  for the light field dynamics. 

\begin{figure}[t]
\centering
\includegraphics[width=\columnwidth]{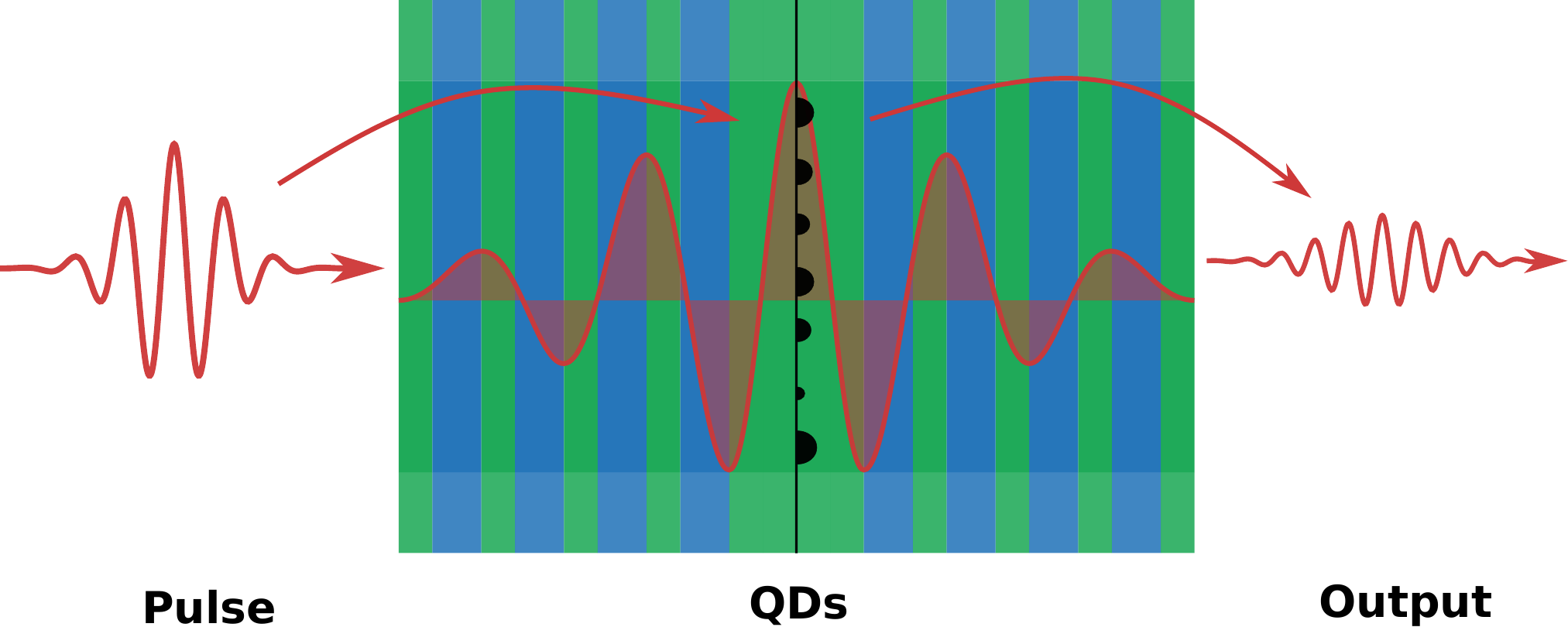}
\caption{Schematic drawing of the simulated situation. A pulse from outside of the cavity excites the cavity mode, which interacts with the QD ensemble inside the layer leading to the emission of an output pulse.}
\label{fig1}
\end{figure}

We take into account the spatial structure along the light propagation direction, perpendicular to a layered structure forming a photonic cavity.
In the cavity a QD ensemble is placed, as schematically sketched in \cref{fig1}.
We are interested in the optical response of this QD ensemble in the cavity upon external driving of the cavity mode.
For this purpose, we assume that an external laser pulse excites the light field and then analyze the output of the system (cf. Fig \ref{fig1}).
We show that, depending on the excitation power, different regimes occur: For low intensities an exciton-polariton like spectrum is observed, while for high amplitudes a spectrum with Rabi splitting emerges.
Interestingly, we find that there is no smooth transition between the regimes but that the transition occurs abruptly at a certain strength of the driving.
Close to the transition nonlinearities strongly affect the dynamics leading to the appearance of a large number of higher harmonics in the spectrum.

We start by performing numerical calculations based on a FDTD method with the embedded few-level systems \cite{chang2004fin,pusch2010con,guazzotti2016dyn,slavcheva2019ult,buschlinger2015lig} and assume them to be identical.
To better interpret our findings, we then show that for the chosen conditions the problem can be reduced to a set of three ordinary non-linear differential equations for the light mode amplitude, the polarization and the occupation of the QD excitons.
These equations can be traced back to a Newton-type equation for an effective particle in a potential, where the shape of the potential depends on the initial excitation of the light field.
Different initial conditions thus lead to different behavior providing an intuitive picture for the transition from the exciton-polariton like dynamics to Rabi oscillations.
Finally we compare the results for identical few-level systems with QD ensembles with Gaussian shaped energetic distributions.
While in the polaritonic regime this leads, as expected, mainly to a broadening of the polariton lines associated with slight shifts \cite{grochol2008mic}, for stronger excitation, but still below the transition to the Rabi oscillation regime, we observe the appearance of additional lines which can be traced back to Rabi oscillations of sub-ensembles of the QDs.

\section{Theoretical background} \label{sec:theory}
Let us start by discussing the photonic structure.
We model a structure, which consists of two Bragg mirrors with $N$ AlAs/GaAs layer pairs on each side surrounding a GaAs cavity of length $l_\te{cav}= \lambda_0/n_\te{GaAs}= 2\pi c /(\omega_0 n_\te{GaAs})$.
Here, $\lambda_0$ and $\omega_0$ are the vacuum wavelength and angular frequency of the cavity mode, respectively, $c=\left( \epsilon_0 \mu_0 \right)^{-1/2}$ is the vacuum speed of light, and $n_\te{GaAs} = 3.535$ the refractive index of GaAs.
Each layer pair consists of a GaAs layer with a width of $l_\te{GaAs} = \lambda_0/(4n_\te{GaAs})$ and an AlAs layer with a width of $l_\te{AlAs} = \lambda_0/(4n_\te{AlAs})$ with the refractive index of AlAs $n_\te{AlAs} = 2.956$.
The cavity frequency is taken to be $\hbar \omega_0 = 1300\, \te{meV}$, which is a typical transition energy for InGaAs QDs \cite{reitzenstein2007ala,bockler2008ele}.
The growth direction of the structure is taken to be the $z$-direction. 

In the center of the cavity, in the plane $z=z_0$, a QD ensemble is placed consisting of $N_\te{QD}$ QDs with transition energies  $\hbar \highi{\omega_x}{n}$, $n=1,\dots,N_\te{QD}$, located at the positions $\vect{r}_n = (x_n, y_n, z_0)$ .
We assume that the spatial distribution of the QDs in the central plane is sufficiently homogeneous such that also the electromagnetic fields can be taken to be homogeneous in $x$- and $y$-direction.
The light field in the structure is described by Maxwell's equations
\begin{subequations}\label{eq:Maxwellvect}
\begin{align}
 \pdv{t} \vect{D}(z,t) &= \nabla \times \vect{H}(z,t)  - \vect{J}_\te{s}(z,t) \label{eq:MaxwellDvect}\\
 \mu_0 \pdv{t} \vect{H}(z,t) &= -\nabla \times \vect{E}(z,t) \label{eq:MaxwellHvect}
\end{align}
\end{subequations}
with the electric field $\vect{E}$, the magnetic field $\vect{H}$, the displacement field $\vect{D}$, and a source current density $\vect{J}_\te{s}$, which will be used to model the external excitation of the system by a short laser pulse.
Note that we will consider only non-magnetic materials.
The displacement field is composed of the electric field $\vect{E}$  and the macroscopic polarization $\vect{P}$ of the system, $\vect{D}(z,t) = \epsilon_0 \vect{E}(z,t) + \vect{P}(z,t)$, where $\vect{P}(z,t)$ consists of the linear polarization of the material of the photonic structure, $\vect{P}_\te{mat}(z,t) = \epsilon_0 \left[n^2(z)-1\right] \vect{E}(z,t)$ with the space-dependent refractive index $n(z)$, and the polarization of the QD ensemble $\mean{\vect{P}}_\te{QD}(z,t)$, averaged over the positions of the QDs.

In general, QDs interacting with a light field which has a frequency close to the lowest exciton transition can be described in terms of a four-level model consisting of the exciton ground state, two single-exciton states with either perpendicular linear or opposite circular polarization, and a biexciton state.
Here, however, we will concentrate on the excitation by circularly polarized light.
In this case the biexciton and the exciton with opposite circular polarization cannot be excited, such that each QD can be reduced to a two-level model consisting of the ground state $ \ket{\highi{g}{n}}$ and the single exciton state $\ket{\highi{x}{n}}$.
The Hamiltonian of the QD ensemble including the light-matter coupling then reads
\begin{align}
  \hat{H} = & \sum_n \hbar \highi{\omega_x}{n} \textstyle{\ketbra{\highi{x}{n}}{\highi{x}{n}} }
    - \vect{E}(z_0,t) \nonumber\\ 
    & \cdot \displaystyle{\sum_n} \biggl( \highi{\vect{M}}{n} \textstyle{\ketbra{\highi{g}{n}}{\highi{x}{n}} } + \vect{M}^{(n)*} \textstyle{\ketbra{\highi{x}{n}}{\highi{g}{n}} } \biggr).
\end{align}
The dipole matrix element for the creation and annihilation of $\sigma_{+}$ ($\sigma_{-}$) polarized excitons is given by $\highi{\vect{M}_{\pm}}{n}= \highi{M}{n} \vect{e}_{\pm}$ with the unit polarization vector $\vect{e}_{\pm}=(\vect{e}_x \pm i \vect{e}_y)/\sqrt{2}$, $\vect{e}_x$ and $\vect{e}_y$ being the unit vectors in $x$- and $y$-direction, $\highi{\vect{M}}{n}$ refers to the dipole matrix element of the excitons with the polarization given by the polarization of the light field, and $\highi{M}{n}$ is taken to be real.  

The microscopic state of the QD $n$ is specified by the microscopic polarization $\highi{p}{n} = \erw{\ketbra{\highi{g}{n}}{\highi{x}{n}}}$ and the exciton occupation $\highi{f}{n} = \erw{\ketbra{\highi{x}{n}}{\highi{x}{n}}}$, which satisfy the Bloch-type equations of motion
\begin{subequations}\label{eq:Bloch}
\begin{align}
 \dv{t} \highi{f}{n} &= -i \frac{\highi{M}{n}\vect{E}(z_0,t)}{\hbar} \cdot \left[ \vect{e}_{\pm} p^{(n)} - \vect{e}_{\mp} p^{(n)*}\right] \label{eq:fn}\\
 \dv{t} \highi{p}{n} &= -i\highi{\omega_x}{n} \highi{p}{n} - i \frac{\highi{M}{n} \vect{E}(z_0,t) \cdot \vect{e}_{\mp}}{\hbar} \left[ 2\highi{f}{n} - 1 \right] \label{eq:pn}\, .
\end{align}
\end{subequations}
The microscopic polarizations of the QDs then give rise to an average macroscopic polarization of the QD ensemble
\begin{align}
 \mean{\vect{P}}_\te{QD} &= \frac{1}{A} \int_A \dd{x} \dd{y} \sum_n \delta(x - x_n) \delta(y-y_n) \delta(z-z_0) \nonumber \\
 & \hspace{0.5cm}\times \int \dd{\omega_x} \delta(\omega_x-\highi{\omega_x}{n}) \left(\highi{\vect{M}}{n} \highi{p}{n} + \vect{M}^{(n)*} p^{(n)*}\right) \nonumber \\
 &= \frac{N_\te{QD}}{A} M \delta(z-z_0) \int \dd{\omega_x} \rho_\te{QD}(\omega_x) \nonumber \\
& \hspace{0.5cm}\times  \left[\vect{e}_{\pm} p(\omega_x,t)+\vect{e}_{\mp} p^*(\omega_x,t)\right] \nonumber \\
  &= \tilde{\vect{P}}(t) \delta(z-z_0), \label{eq:macr_P}
\end{align}
where 
\begin{align}
\tilde{\vect{P}}(t) &=  \frac{MN_\te{QD}}{A} \int \dd{\omega_x} \rho_\te{QD}(\omega_x) \nonumber \\
& \quad \times \left[\vect{e}_{\pm} p(\omega_x,t) + \vect{e}_{\mp} p^*(\omega_x,t)\right] \label{eq:pol_vect}
\end{align}
with the normalization area $A$ and the number $N_\te{QD}$ of QDs in the area $A$. Here, 
\begin{align}
\rho_\te{QD}(\omega_x)= \frac{1}{M N_\te{QD}} \sum_n \delta(\omega_x-\highi{\omega_x}{n}) \highi{M}{n}
\end{align} 
is a normalized distribution function of QD transition energies $\omega_x$ weighted by the dipole moments  and $p(\omega_x)$ is the microscopic polarization of the QDs with this transition energy.
Correspondingly, $f(\omega_x)$ is the exciton occupation of these QDs.
The average dipole moment is defined by
\begin{align}
M=\frac{1}{ N_\te{QD}} \sum_n \highi{M}{n}.
\end{align}

As mentioned above, in the following we will restrict ourselves to circularly polarized light.
However, since we use real electromagnetic fields some care has to be taken in identifying the respective contributions.
For a $\sigma_{+}$ ($\sigma_{-}$) circularly polarized light field with central frequency $\omega_0$ traveling in $z$-direction, the $y$-component of the electric field follows (precedes) the $x$-component by a quarter period, i.e., we have
\begin{align}
\vect{E}(z,t) = \frac{1}{\sqrt{2}} \left[ E\left(z,t\right) \vect{e}_x \pm E\left(z,t-\frac{\pi}{2\omega_0}\right) \vect{e}_y \right]. \label{eq:E_circ}
\end{align}
The magnetic field then satisfies
\begin{align}
\vect{H}(z,t) = \frac{1}{\sqrt{2}} \left[ H\left(z,t\right) \vect{e}_y \mp H\left(z,t-\frac{\pi}{2\omega_0}\right) \vect{e}_x \right].
\end{align}
In order to excite such a field, also the source current density in \cref{eq:MaxwellDvect} has to be of the same structure, i.e.,
\begin{align}
\vect{J}_\te{s}(z,t) = \frac{1}{\sqrt{2}} \left[ J_\te{s}\left(z,t\right) \vect{e}_x \pm J_\te{s}\left(z,t-\frac{\pi}{2\omega_0}\right) \vect{e}_y \right]. \label{eq:J_circ}
\end{align}
As is shown in the appendix \ref{app:circular}, as long as we restrict ourselves to light field amplitudes which vary on a time scale much longer than the oscillation period $2\pi/\omega_0$, the electric field in \cref{eq:E_circ} can be rewritten as 
\begin{align}
\vect{E}(z,t) = \frac{1}{2} \left[ \vect{e}_{\pm} \tilde{E}\left(z,t\right) e^{-i\omega_0 t} + \vect{e}_{\mp} \tilde{E}^*\left(z,t\right) e^{i\omega_0 t} \right]. \label{eq:E_circ2}
\end{align}
with the slowly varying complex amplitude $\tilde{E}$.
Inserting this field in \cref{eq:Bloch} and using $\vect{e}_{\pm} \cdot \vect{e}_{\mp}=1$ and $\vect{e}_{\pm} \cdot \vect{e}_{\pm}=0$, we see that indeed a $\sigma_{+}$ ($\sigma_{-}$) circularly polarized light field only excites the $\sigma_{+}$ ($\sigma_{-}$) circularly polarized exciton, except for negligible contributions resulting from counter-rotating terms $\sim \exp[\pm 2 i\omega_0 t]$.
Following again the derivation in the appendix \ref{app:circular}, we find that also the polarization in \cref{eq:pol_vect} can be separated into $x$- and $y$-components according to 
\begin{align}
\tilde{\vect{P}}(t) &= \frac{1}{\sqrt{2}}  \left[ \tilde{P}(t) \vect{e}_x \pm \tilde{P}\left( t-\frac{\pi}{2\omega_x} \right) \vect{e}_y \right],
\end{align}
with
\begin{align}
\tilde{P}(t) =  \frac{MN_\te{QD}}{A} \int \dd{\omega_x} \rho_\te{QD}(\omega_x) \left[p(\omega_x,t) + p^*(\omega_x,t)\right] \label{eq:macr_Pmean}.
\end{align}
This shows that indeed for all vector fields entering Maxwell's equations (\ref{eq:Maxwellvect}) the $y$-component agrees with the $x$-component shifted by a quarter period.
Therefore, it is sufficient to solve the equations for the $x$-component of the electric and the $y$-component of the magnetic field, resulting in the final set of equations of motion
\begin{subequations}\label{eq:eom}
\begin{align}
 \pdv{t} E(z,t) &= - \frac{1}{n^2(z) \epsilon_0} \left[ \pdv{z} H(z,t) \right. \nonumber \\
 & \hspace{0.5cm}\left. {} + \delta(z-z_0) \pdv{t} \tilde{P}(t) -  J_\te{s}(z,t) \right],\label{eq:1D_maxwell_E}\\
 \pdv{t} H(z,t) &= -\frac{1}{\mu_0} \pdv{z} E(z,t), \label{eq:1D_maxwell_H}\\
 \dv{t} f(\omega_x,t) &= 2\frac{ME(z_0,t)}{\hbar} \Im\left(p(\omega_x,t)\right), \label{eq:f}\\
 \dv{t} p(\omega_x,t) &= -i\omega_x p(\omega_x,t) - i \frac{ME(z_0,t)}{\hbar} \left[ 2f(\omega_x,t) - 1 \right]. \label{eq:p}
\end{align}
\end{subequations}
\Cref{eq:eom,eq:macr_Pmean} are solved numerically.
For \cref{eq:f,eq:p} we use a standard fourth order Runge-Kutta method and \cref{eq:1D_maxwell_E,eq:1D_maxwell_H} are implemented using a FDTD method in one spatial dimension \cite{taflove2005com}.
The system is excited at a position $z_1$ outside of the cavity with a Gaussian current density
\begin{align}
J_\te{s}(z,t)=J \delta\left( z-z_1 \right) \exp\left[-\frac{4t^2 \ln(2)}{\tau^2}\right] \cos(\omega_0 t)
\end{align}
with a full width at half maximum (FWHM) of $\tau=200$~fs.
Reflections from the boundaries of the simulation region are avoided by using perfectly matched absorbing boundary layers \cite{taflove2005com}.

The electric field in \cref{eq:1D_maxwell_E} has two driving terms, one involving the current density $J_\te{s}$ and the other the QD polarization $\tilde{P}(t)$.
Due to the linearity of \cref{eq:1D_maxwell_E,eq:1D_maxwell_E} the electric (and also the magnetic field) can be separated into two contributions, 
\begin{align}
E(z,t) = E_\te{ext}(z,t) + E_\te{ind}(z,t) \label{eq:E_ext_ind}
\end{align}
with the external field $E_\te{ext}(z,t)$ driven only by the current density and the induced field $E_\te{ind}(z,t)$ driven only by the polarization.
We will come back to this separation below when discussing the results.

In the following sections we will first concentrate on the case of identical QDs in resonance with the light field, i.e., a sharp distribution
\begin{align}
\rho_\te{QD}(\omega_x) = \delta(\omega_x - \omega_0)
\end{align}
In Sec.~\ref{sec:distribution} we will then analyze the influence of a non-vanishing width of the QD distribution on the results.

\section{Characterization of the system}
\begin{figure}[t]
\centering
\includegraphics[width=\columnwidth]{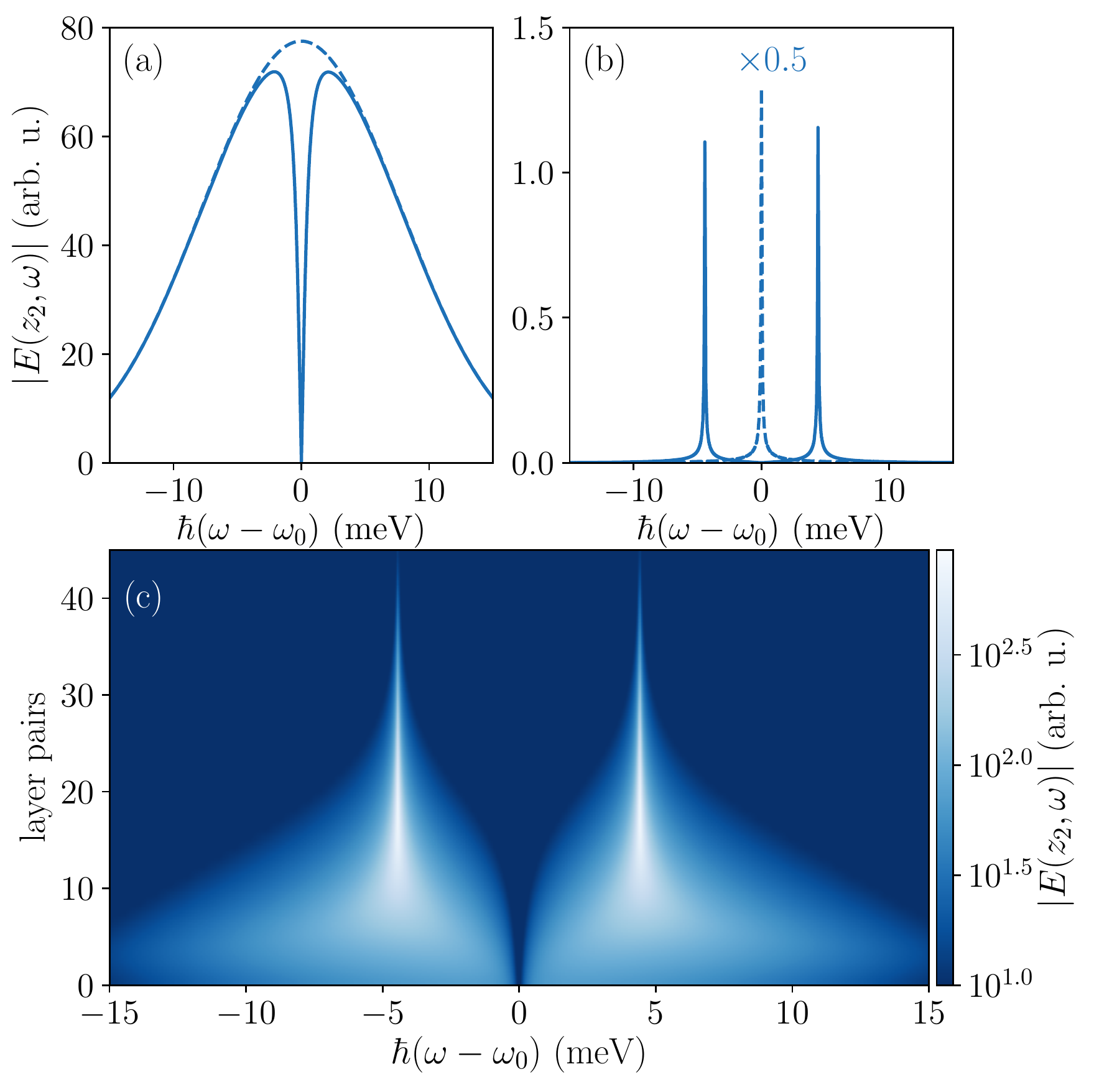}
\caption{Linear transmission spectra for excitations with low amplitude (a) without photonic structure and (b) for a cavity with $N=30$ Bragg layer pairs on each side.
The dashed lines correspond to the spectra without QDs, the solid lines show the spectra with QD ensemble.
(c) Transmission spectrum for increasing number of layer pairs of the Bragg mirror at each side of the cavity. The spectra (b) and (c) are calculated using a small exponential damping for better visualization.}
\label{fig2}
\end{figure}

Before considering the dynamics of the system, we briefly characterize the photonic cavity with the QDs by analyzing its transmission spectrum.
For this purpose, we excite the system with a small amplitude $J$ at the position $z_1$ on the left side and calculate the Fourier transformation of the electric field after passing the system, i.e., at a position $z_2$ on the right side.
Without a cavity, the pulse has a Gaussian spectral shape.
When the pulse  interacts with the QDs, a dip in the spectrum is found as displayed in Fig.~\ref{fig2} (a).
The width of this absorption dip depends on the lifetime of the excited state.
The decay is an effect of the radiative interaction with the field induced by the QD polarization \cite{kirakoch2012sem,taniyama2019sim} and here depends on $M \frac{N_\te{QD}}{A}$ (see \cref{eq:macr_Pmean}.) 

Now we put a cavity around the QD ensemble.
\Cref{fig2} (b) shows the linear transmission spectrum of the QD ensemble in a cavity with Bragg mirrors of $30$ layer pairs on each side.
Without QDs (dashed line) we see a sharp peak of the cavity mode at $\omega = \omega_0$.
With QDs (solid line), this peak splits up in two separate peaks.
The splitting occurs due to the strong light confinement and enhanced light-matter interaction.
Accordingly, the double peak only exists if the confinement is strong, which depends on the number of Bragg layers.
This is demonstrated in \cref{fig2} (c), showing the transmission spectrum with QDs for different number of layer pairs on each side.
For small $N$ there is a broad spectrum with a dip as found in \cref{fig2} (a).
For values $N>10$ the formation of the coupled QD-light state can be observed. We emphasize that the splitting is independent of the number of layers, once this state exists.

We have checked that the splitting does not depend on the amplitude as long as we stay in the limit of small amplitudes, but is proportional to $\sqrt{N_\te{QD}/A}$ and $M$.
We will come back to this dependence below in Sec.~\ref{sec:analytical}.
We note that these peaks exhibit an anti-crossing behavior when tuning the transition frequency of the QDs through resonance of the cavity (not shown).
These properties are typical for exciton-polaritons \cite{laucht2009dep}.

\section{Transition from low to high amplitudes}\label{sec:results_num}
\begin{figure}[t]
\centering
\includegraphics[width=\columnwidth]{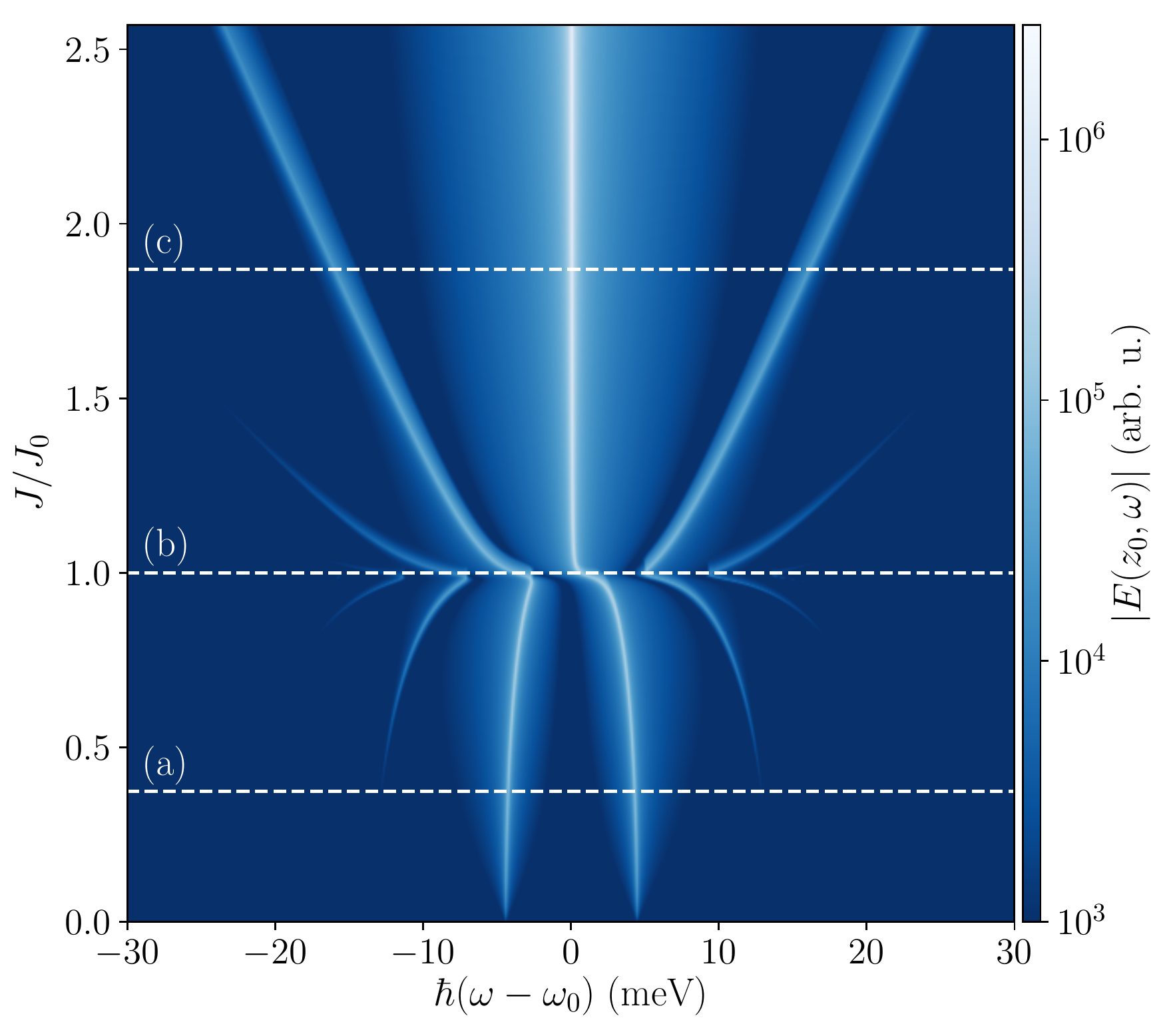}
\caption{Spectrum of the field inside the cavity for varying amplitude $J$ of the exciting pulse, given in units of the threshold amplitude $J_0$.
The marked horizontal lines correspond to the cases shown in \cref{fig4}.}
\label{fig3}
\end{figure}

Now we study the spectrum, when increasing the amplitude $J$ of the external current.
The resulting spectra are shown in \cref{fig3}, where we fixed the number of layer pairs to $N=30$.
Note that here we have plotted the spectra of the field at the position of the QDs.
We checked that these agree qualitatively with the transmission spectra. 

The lower part of this figure corresponds to the excitation with a small pulse amplitude and shows a double peak structure as discussed in the previous section.
For low amplitudes the splitting is essentially independent of the amplitude.
When increasing the amplitude further the splitting slightly reduces and additional side bands are formed.
These are nonlinear effects of the QD-cavity coupling, which result from the re-interaction with the induced light field. 

At a certain amplitude $J_0$ of the driving current density a sudden transition takes place (see line (b) in \cref{fig3}) and the behavior of the spectrum changes qualitatively.
For higher pulse amplitudes we now obtain three peaks in the spectrum with a main peak at $\omega = \omega_0$ and two satellite peaks.
The main peak stays at $\omega = \omega_0$ when increasing the amplitude, while the side peaks change their position with increasing $J$.
For sufficiently strong excitation the splitting of the side peaks grows linearly with the $J$. 

\begin{figure}[t]
\centering
\includegraphics[width=\columnwidth]{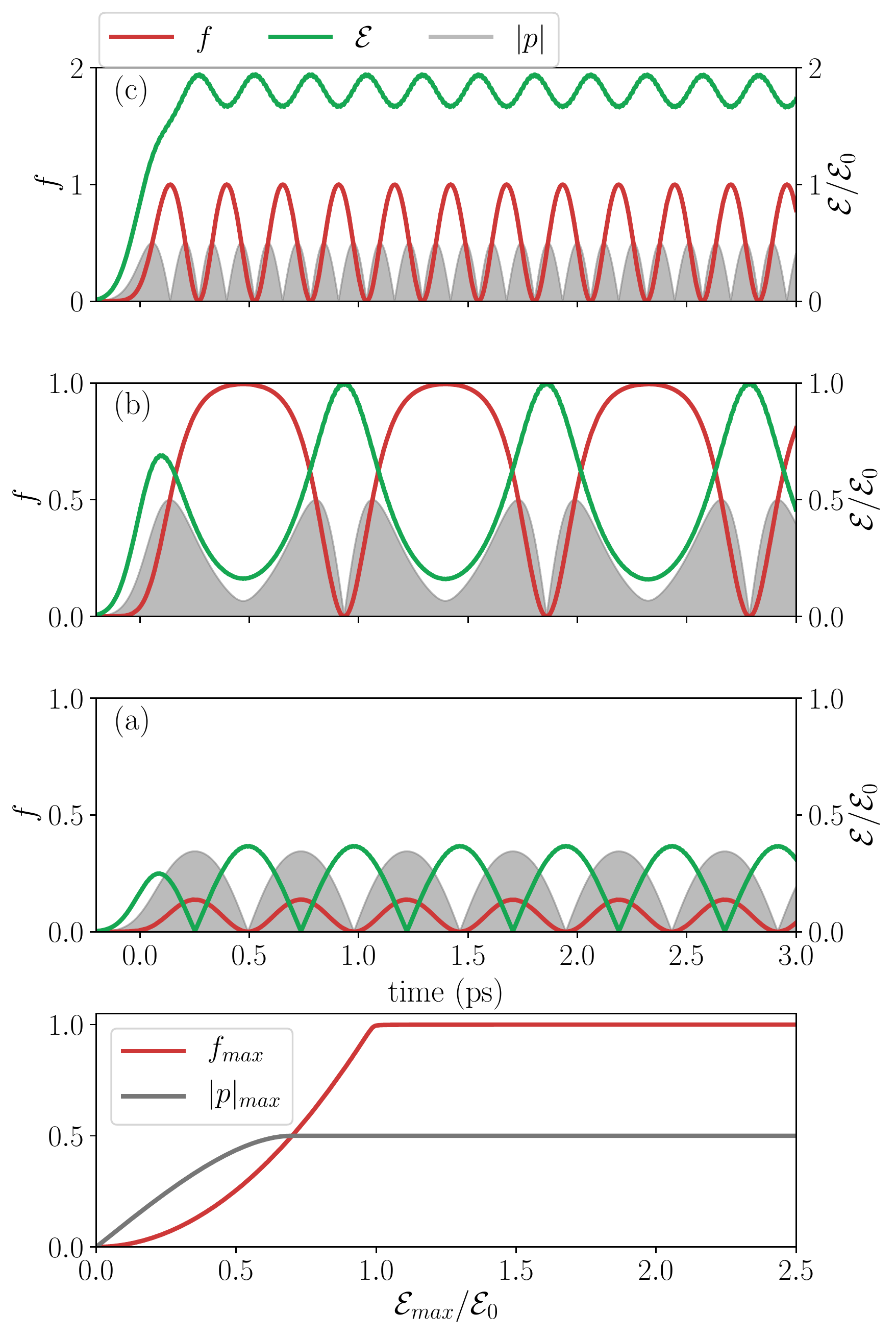}
\caption{Occupation (red lines), polarization (gray area), and light field (green lines) dynamics inside a cavity with $N=30$ Bragg layer pairs on each side for different driving pulse amplitudes $J$ corresponding to the cases marked by (a), (b), and (c) in \cref{fig3}.
Bottom panel: maximum occupation and polarization depending on the maximum field amplitude $\Epsilon_\te{max}$ at the position of the QDs.
$\Epsilon_0$ denotes the maximum field amplitude at the transition.}
\label{fig4}
\end{figure}

To understand the difference in the two regimes,  in \cref{fig4} we look at the dynamics of the occupation $f$ (red line), the amplitude $\Epsilon=|\tilde{E}(z_0,t)|$ of the slowly varying part of the electric field at the position of the QDs (green line) and the absolute value of the polarization $|p|$ (gray area) for three different cases: (a) low amplitude with a double peak, (b) at the transition and (c) in the high intensity regime with three peaks.
These cases are marked by white lines in \cref{fig3}. The field amplitude is normalized to $\Epsilon_0$ being the maximal electric field amplitude at the transition (see line (b) in \cref{fig3}).

All dynamics start with a switch on of the electric field around $t=0$ due to the pulsed current density $J_\te{s}$, followed by an oscillatory behavior.
At the switch on, first the electric field builds up, which then induces a polarization, and finally an occupation of the QD system is created.
This can also be seen in \cref{eq:f,eq:p} where the electric field drives the polarization and the polarization drives the occupation.

For low amplitudes, shown in \cref{fig4} (a), we find oscillations for all three quantities.
The reason for this is that the electric field $E(z_0,t)$ builds up a polarization.
The polarization, in turn, creates an induced field $E_\te{ind}(z_0,t)$ which is opposite to the external field $E_\te{ext}(z_0,t)$, such that the amplitude of the total field $\Epsilon$ decreases (see \cref{eq:E_ext_ind}).
When the polarization reaches a maximum $\Epsilon$ vanishes.
Then the polarization acts as a source for the electric field.
Due to the strong confinement, the emitted light is trapped in the cavity and builds up the cavity mode again.
This induces the oscillations between $|p|$ and $\Epsilon$.
The occupation follows the polarization.
In this regime of weak excitations none of the quantities reaches one.
The occupation is quadratic in the polarization and thus also in the electric field, therefore it remains small at all times.
Such a periodic oscillation of the energy between electric field and polarization is characteristic for polaritonic dynamics.
The two spectral lines reflect the lower and the upper polaritonic branch.

The dynamics for high amplitudes is shown in \cref{fig4}~(c).
Also here we observe - after an initial transient - oscillations of all three quantities, field, polarization and occupation.
However, now the occupation oscillates between 0 and 1 and it is out of phase with the polarization.
This is a characteristic behavior for Rabi oscillations of a two-level system.
The amplitude of the electric field exhibits weak oscillations around a non-zero value, in contrast to the weak driving it never reaches zero.
This non-zero mean value is the reason for the central peak at $\omega_0$.
For high amplitudes the induced field is much smaller than the external one, because the driving term $\mean{P}_\te{QD}$ from \cref{eq:macr_Pmean} is limited by the microscopic polarization which, in turn is limited by the number of QDs.
Therefore $E_\te{ext}$ is the main contribution to the QD dynamics and approximately acts as a continuous wave excitation for the two-level system.
From the classical Rabi model one expects Rabi oscillations with Rabi frequency $\Omega_\text{R} = \frac{\left|\tilde{E}_\te{ext} M\right|}{\hbar}$, where $\tilde{E}_\te{ext}$ denotes the amplitude of the external field.
In the dressed state picture this corresponds to two states, which are separated by the Rabi energy $\hbar\Omega_\text{R}$, when driven resonantly \cite{gerry2005int}.
The oscillating polarization of the Rabi oscillations creates an induced field oscillating with the same frequency which contributes to the total field and leads to the frequency contributions at $\omega = \omega_0 \pm \Omega_\text{R}$.
This is the origin of the satellite peaks in the upper regime of \cref{fig3}. 

The three-peak structure of the spectrum with a central peak and two satellite peaks shifted by $\pm\Omega_\text{R}$ reminds one of the Mollow triplet seen in the resonance fluorescence of a two-level system driven by a classical light field \cite{mollow1969pow}. Indeed, it has the same physical origin. However, while the Mollow triplet with its characteristic intensity ratio between the three peaks is a quantum optical effect seen in the power spectrum of a resonantly scattered additional light field, here, the peaks appear in the spectrum of the driving field itself, which is modified by the field induced self-consistently by the QD polarization. The peaks have no fixed intensity ratio, instead the relative intensity of the central peak increases with increasing driving because, as discussed above, the induced field is limited by the number of QDs.

The oscillating part of the electric field couples back to the dynamics of occupation and polarization.
Due to the non-linearity of these equations this gives rise to the creation of higher harmonics of the Rabi frequency.
They are particularly pronounced slightly above the threshold field $\Epsilon_0$ because here the induced field is of the same order as the external field, while its relative importance decreases with increasing external driving.

Finally we look at the transition region where $J \approx J_0$.
When increasing the pulse amplitude, the occupation increases likewise up to the point where the occupation almost reaches one.
The dynamics of field, occupation, and polarization in this regime are shown in \cref{fig4} (b).
We still observe an oscillatory behavior in all three quantities, but the oscillations strongly deviate from a sinusoidal shape.
The occupation now reaches its maximal value of unity, and around this maximum plateaus show up.
Furthermore, $\Epsilon$ does not reach zero like in the low driving case, but gets close to zero when $f$ reaches its maximum.
If the total field is very small, the actual Rabi frequency is small too, and the occupation only varies slowly.
This results in the formation of the plateaus as seen in \cref{fig4} (b) and a longer oscillation periodicity resulting in the reduction of the splitting in the spectrum as well as the appearance of higher harmonics at line (b) in \cref{fig3}.

The bottom panel in \cref{fig4} displays the maximum values of the occupation and the polarization as a function of the maximum value of the field. At low fields we see the linear increase of the polarization and the quadratic increase of the occupation. We also clearly see that the transition between the different regimes is reached when the maximum of the occupation reaches unity.

Let us briefly compare our results for a QD ensemble in the plane $z=z_0$ with the case of a single QD in a microcavity.
In that case a quantized description of the light field is necessary, which is achieved by the Jaynes-Cummings model.
In our scenario, the cavity field is driven by an external laser pulse, which leads to the creation of a coherent state of the cavity with mean photon number depending on the driving strength.
In the low driving limit the mean photon number is much smaller than one, such that the field consists of a superposition of the zero-photon and the one-photon state with negligible contributions from higher photon states.
Since the QD is initially in its ground state, there are no dynamics in the zero-photon subspace, while in the one-photon subspace Rabi oscillations with the vacuum Rabi splitting set in.
This corresponds to the polariton regime with the vacuum Rabi splitting being the polariton splitting.
For very strong driving, on the other hand, a coherent state with a high mean photon number and thus small relative uncertainty in the photon number is generated.
This leads approximately to Rabi oscillations with the Rabi frequency corresponding to the mean photon number, giving rise to the three peak structure as seen in our case, too.
However, in the transition region the system is driven into the well-known collapse-and-revival regime because many different photon numbers and consequently different Rabi frequencies with in general irrational ratios are present.
This is in clear contrast to the findings in our case of a sharp transition between the two regimes and periodic oscillations at all driving amplitudes, however with strongly anharmonic shape in the region close to the transition.

Indeed, both the polaritonic and the Rabi oscillation have been seen in structures with a single QD.
The polariton-like behavior in the strong coupling limit \cite{reithmaier2004str} and the Mollow-triplet \cite{ulrich2011dep,ulhaq2013det,reiter2017tim,roy2012pol}, similar to the three-peak structure in the Rabi model, have been independently studied and have recently been connected in an experiment, where a QD in a micropillar cavity is optically excited from the side \cite{hopfmann2017tra}.
There the transition between the purely quantum model of the Jaynes-Cummings ladder with the semiclassical Autler-Townes ladder was investigated using a Jaynes-Cummings model showing a transition region between the two regimes, but no sharp transition.

In the next section we will show that our system can be mapped to the dynamics of a particle in an effective, anharmonic potential, which will allow us to obtain a deeper insight in the different regimes and the sharp transition between them.

\section{Analytical Model}\label{sec:analytical}
The results in the previous section \ref{sec:results_num} have been obtained for a cavity with a high quality factor, which means that cavity losses are negligible on the considered time scales.
In fact, from the FDTD simulation we can extract for the cavity with 30 layer pairs in each Bragg mirror a Q factor of 870,000 corresponding to a photon lifetime of 880 ps, which is much longer than the typical time scales of one or a few picoseconds (see \cref{fig4}).
In this case, the light field in the cavity can be expanded in cavity modes.
We want to remark that this approach can be extended to leaky cavities by using an expansion into quasinormal modes of the cavity \cite{ching1998qua}.

Combining Maxwells equations \cref{eq:1D_maxwell_E,eq:1D_maxwell_H} we obtain the driven wave equation for the electric field 
\begin{align}  \label{eq:a_waveequation}
 \frac{-c^2}{n^2(z)} \pdv[2]{z} E(z, t) + \pdv[2]{t} E(z, t) = \frac{-1}{n^2(z)\epsilon_0} \pdv[2]{t} \mean{P}_\te{QD} \,.
\end{align}
We have omitted the source current density because here we will replace the pulsed excitation by chosing an initial value of the electric field.
For the homogeneous part of the wave equation we can define the eigenvalue problem 
\begin{align}
 \hat{\Phi} u_m = \omega_m^2 u_m \quad \text{with} \quad \hat{\Phi} =- \frac{c^2}{n^2(z)} \dv[2]{z}\label{eq:a_def_u} \, ,
\end{align}
where we have defined the differential operator $\hat{\Phi}$ with the eigenvalues $\omega_m^2$ and eigenfunctions $u_m$.
Note that $\hat{\Phi}$ is not self-adjoint.
However, by taking the derivative of the eigenvalue problem we obtain
\begin{align}
 \hat{\tilde{\Phi}} \tilde{u}_m = \omega_m^2 \tilde{u}_m 
 \end{align}
 with
 \begin{align}
 \hat{\tilde{\Phi}} =- \dv{z} \frac{c^2}{n^2(z)} \dv{z}\quad \te{and} \quad \tilde{u}_m = \dv{u_m}{z}.
\end{align}
Now the operator $\hat{\tilde{\Phi}}$ is self-adjoint, which ensures that the functions $\tilde{u}_m(z)$ constitute a complete orthonormal set of eigenfunctions and the eigenvalues $\omega_m^2$ are real.
Using a partial integration, the orthonormality condition can be rewritten as
\begin{align}
\delta_{km}&= \int \tilde{u}_k^* \tilde{u}_m\, \dd{z} = \int \tilde{u}_k^* \dv{u_m}{z} \, \dd{z} \nonumber \\ &= - \int \dv{\tilde{u}_k^*}{z} u_m(z)\, \dd{z} 
= - \int \dv[2]{u_k^*}{z} u_m(z)\, \dd{z},
\end{align}
showing that the function $v_k^*(z) = - d^2u_k^*(z)/dz^2$ is the orthogonal eigenfunction to $u_k(z)$.
Here we have assumed either periodic boundary conditions or vanishing mode functions far away from the cavity, which is well satisfied due to the high quality factor of the cavity.
Then we can expand the electric field into the eigenmodes $u_m(z)$,
\begin{align}
 E(z, t) = \sum_m E_m(t) u_m(z) \, .\label{eq:a_eigenmodeexpansion}
\end{align}
Using \cref{eq:a_eigenmodeexpansion} and the orthogonality of $v_k$ and $u_m$, we can rewrite the wave equation \cref{eq:a_waveequation} as
\begin{align}
 \dv[2]{t} E_k(t) + \omega_k^2 E_k(t) = -\frac{v_k^*(z_0)}{n^2(z_0)\epsilon_0} \dv[2]{t} \tilde{P}(t)\, ,
\end{align}
where we inserted the polarization $\mean{P}_\te{QD} = \tilde{P}(t) \delta(z-z_0)$.

Assuming a sufficiently sharp QD ensemble with transition energies close to the frequency of one of the cavity modes and width much smaller than the mode separation to the other cavity modes, only a single cavity mode $E_c$ with frequency $\omega_c= \omega_0$ is effectively coupled to the QDs.
Defining 
\begin{align}
E_c(t) = \frac{1}{2} \left[\tilde{E}_c(t) e^{-i\omega_0 t}+\tilde{E}^*_c(t) e^{i\omega_0 t}\right]
\end{align}
and $p(\omega_x, t) = \tilde{p}(\omega_x, t) e^{-i\omega_0 t}$ we apply both the rotating-wave approximation (RWA) and the slowly-varying-amplitude approximation.
Within these approximations we obtain the equation of motion for the field amplitude
\begin{subequations} \label{eq:eom_analytical}
\begin{align}
 \dv{t} \tilde{E}_c(t) &= i \tilde{M} \lambda \int \dd{\omega_x} \rho_\te{QD}(\omega_x) \tilde{p}(\omega_x, t) \label{eq:a_E} \,, 
\end{align}
which is complemented by the equations of motion for $f(\omega_x, t)$ and $\tilde{p}(\omega_x, t)$
\begin{align}
 \dv{t} f(\omega_x, t) &= -i \tilde{M} \left[\tilde{E}_c^*(t) \tilde{p}(\omega_x, t) - \tilde{E}_c(t) \tilde{p}^*(\omega_x, t) \right] \label{eq:a_f}\\
 \dv{t} \tilde{p}(\omega_x, t) &= -i(\omega_x - \omega_0)\tilde{p}(\omega_x, t) \nonumber \\
    &-i\tilde{M} \tilde{E}_c(t) \left[ 2f(\omega_x, t) - 1 \right] \label{eq:a_p}\, .
\end{align}
\end{subequations}
Here we have introduced the abbreviations $\tilde{M} = M u_c(z_0)/(2\hbar)$ and $\lambda = \frac{v_c^*(z_0)}{u_c(z_0)}\frac{N_\te{QD}}{A} \frac{2\hbar \omega_0}{n^2(z_0)\epsilon_0}$.
The constants defined via material parameters are given in Sec.~\ref{sec:theory}, while $u_c(z)$ and $v_c(z)$ can be either calculated using FDTD or by solving \cref{eq:a_def_u} for the normal modes.
Equations (\ref{eq:eom_analytical}) are a set of coupled nonlinear equations of motion, which define the dynamics of $\tilde{E}_c(t)$, $f(\omega_x, t)$ and $\tilde{p}(\omega_x, t)$ by their initial values.
In this section, we consider $\rho_\te{QD}(\omega_x) = \delta(\omega_x - \omega_0)$ and the initial values at $t=0$ are $\tilde{E}_c(0) = \tilde{E}_0$, $f(\omega_x, 0) = 0$ and $\tilde{p}(\omega_x, 0) = 0$ in accordance with our numerical simulations.
Indeed, when we numerically solve \cref{eq:eom_analytical} for different initial values $\tilde{E}_0$, we find basically the same dynamical and spectral behavior of this system as described in Sec.~\ref{sec:results_num} when tuning the pulse amplitude.
Therefore we do not repeat the corresponding figures.

The set of equations can be further simplified by introducing nondimensionalized quantities
\begin{subequations}
\begin{align}
	\alpha(\tau) &= \frac{\tilde{M} \tilde{E}_c(\tau)}{\sqrt{\tilde{M}^2 \lambda}},~~ 
	&\beta(\tau) &= (2f(\tau)-1), \\
	\gamma(\tau) &= ip(\tau),~~
	&\tau &= \sqrt{\tilde{M}^2 \lambda}\, t\, .
\end{align}
\end{subequations}
Thereby we assumed that $\tilde{E}_c$ is real, such that the polarization is purely imaginary and $\alpha$, $\beta$, and $\gamma$ are real functions.
Then the new equations of motion read
\begin{subequations}
\label{eq:anal_eom}
 \begin{align}
 \dv{\tau} \alpha(\tau) &= \gamma(\tau) \label{eq:a_alpha} \\
 \dv{\tau} \beta(\tau) &= -4\alpha(\tau) \gamma(\tau) \label{eq:a_beta} \\
 \dv{\tau} \gamma(\tau) &= \alpha(\tau) \beta(\tau) \label{eq:a_gamma}.
\end{align}
\end{subequations}
It is interesting to note that we now have eliminated all system parameters in the equations of motion, which makes them a prototypical example of coupled, nonlinear differential equations.
The initial values at $\tau=0$ are given by $\alpha(\tau=0) = \alpha_0$, $\beta(\tau=0) = \beta_0=-1$ and $\gamma(\tau=0)=\gamma_0=0$.
Here $\alpha$ is a measure for the electric field strength at the location of the QDs.

Let us start by discussing the dynamics in the limiting cases of a very low and a very high initial value of $\alpha$.
For low $\alpha_0$, we find that $2f = \beta - \beta_0$ is quadratic in the electric field $E \sim \alpha$. Restricting ourselves to the linear regime, we neglect the differential equation for $\beta$ and set $\beta = \beta_0 = -1$, resulting in two equations for $\alpha$ and $\gamma$ \cref{eq:a_alpha,eq:a_gamma}, which can be reduced to
\begin{align}
 \dv[2]{\tau} \alpha(\tau) &= -\alpha(\tau) \nonumber \,. 
\end{align}
This is the equation of a harmonic oscillator with a constant frequency of unity. In the dimensionalized problem, the oscillation frequency is $\Omega_p = \pm \tilde{M} \sqrt{\lambda}$.
This is in agreement with the numerical findings in \Cref{sec:results_num}.
Indeed, this corresponds to the splitting in two exciton-polariton states as seen in the FDTD calculations (lower part of \cref{fig3}).

In the case of a high $\alpha_0$ the influence of the macroscopic polarization is small, such that we approximate $\dv{\tau} \alpha \to 0$ and $\alpha \to \alpha_0$.
Then \cref{eq:a_beta,eq:a_gamma} can be written as a harmonic oscillator
\begin{align}
 \dv[2]{\tau} \beta(\tau) &= -4 \alpha_0 \beta(\tau) \nonumber \,. 
\end{align}
with frequency $\Omega_\te{R} = \pm 2 \abs{\alpha_0}$ or transforming back with $ \Omega_\te{R}= \pm 2 \abs{\tilde{M}\tilde{E}}$.
This oscillation weakly couples back to $\alpha$ leading to the Rabi splitting, which has been found in the FDTD calculations in the upper part of \cref{fig3}.

The analytical model allows us to furthermore understand the qualitative difference in the two regimes and also the transition region of \cref{fig3}.
For this purpose, we first combine \cref{eq:a_alpha} and \cref{eq:a_beta}
\begin{align}
 \dv{\tau} \beta(\tau) &= -4\alpha(\tau) \gamma(\tau) = -4\alpha(\tau) \dv{\tau}\alpha(\tau) \nonumber 
\end{align} 
and then solve by integration
\begin{align}
 \Rightarrow \beta(\tau) &= -1 - 4 \int \limits_{0}^{\tau} \dd{t'} \alpha(t') \dv{t'}\alpha(t') \nonumber \\
    &= -1 - 2 \alpha^2(\tau) + 2 \alpha_0^2 \label{eq:beta_int} \, ,
\end{align}
where we made use of the initial condition $\beta_0 = -1$.
Next we combine \cref{eq:a_alpha} and \cref{eq:a_gamma} to
\begin{align}
 \dv[2]{\tau} \alpha(\tau) = \dv{\tau} \gamma(\tau) = \alpha(\tau) \beta(\tau) \nonumber 
\end{align}
and use \cref{eq:beta_int} to obtain
\begin{align}
\dv[2]{\tau} \alpha(\tau) = -2\alpha(\tau)^3 + \left(2 \alpha_0^2 - 1 \right) \alpha(\tau) \, .
\end{align}
This is a Newton type equation of motion $\dv[2]{\tau} \alpha = -\pdv{\alpha} V(\alpha)$ for a particle moving in a potential $V(\alpha)$ defined as 
\begin{align}
 V(\alpha) &= \frac{1}{2} \left( \alpha^4 - \left( 2\alpha_0^2 - 1 \right) \alpha^2 \right). \label{eq:a_potential}
\end{align}
We stress that the potential $V$ itself already depends on the initial values of the system.
$V(x)=ax^4+bx^2$ is a typical potential leading to phase transitions when the single minimum splits into a double minimum.
However, in our case, we have no damping which would drive the system into one of the minima.
Instead, the initial condition determines both the potential shape and the starting point of the dynamics and thereby the dynamical behavior.
We have solved this equation numerically and extracted the spectrum of the electric field, i.e., $|\alpha(\omega_\tau)|$.
The resulting figure is indistinguishable from the one shown in \cref{fig3}, confirming that the conditions for the reduction to the single cavity mode model and the various approximations discussed above are indeed almost perfectly satisfied. 

\begin{figure}[t]
\centering
\includegraphics[width=\columnwidth]{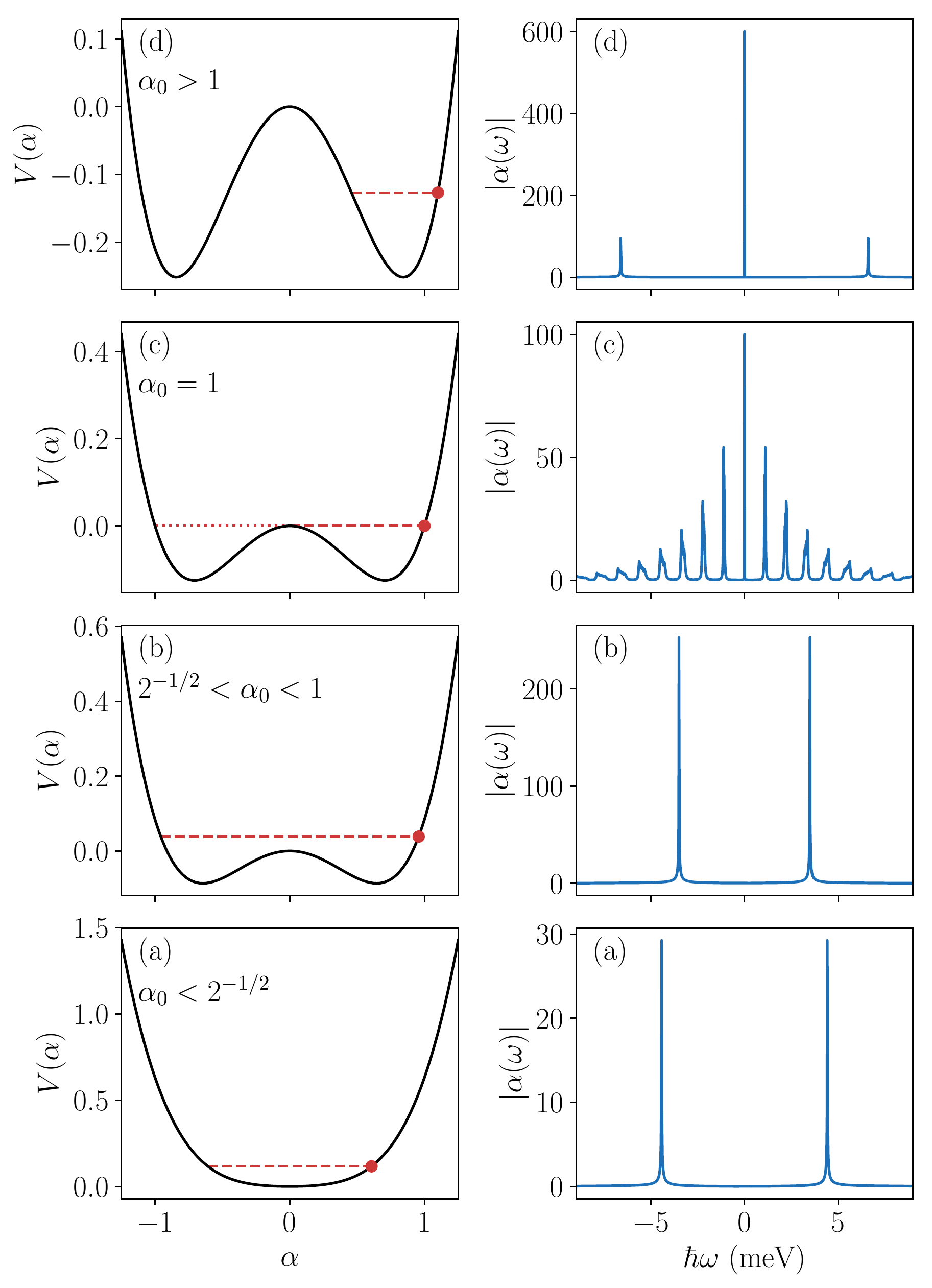}
\caption{Left: Potential $V$ for different initial values (red dots) of $\alpha_0$. The red dashed line marks the allowed values  for the dynamics.
Right: Resulting spectra of the corresponding dynamics at the values (a) $\alpha_0=0.1$, (b) $\alpha_0=0.8$, (c) $\alpha_0\approx 1$, (d) $\alpha_0=1.1$.}
\label{fig5}
\end{figure}

To get a deeper insight, in \cref{fig5}~(left) we plot four distinct cases of the potential for different values of $\alpha_0$, the right column shows the corresponding spectrum $|\alpha(\omega)|$.
Note that because $\alpha$ describes the slowly varying envelope of the electric field the frequency axis is shifted such that the cavity mode now corresponds to the frequency zero.
The ``particle'' starts with zero velocity at the initial position $\alpha_0$ marked by the red dot on the potential curve.
As in classical physics, every point of the graph below the red line can be reached during the dynamics.
In the case of a low initial field with $\alpha_0 < {2}^{-1/2}$ (\cref{fig5}~(a)), the potential has only one minimum and we have an oscillation between $-\alpha_0$ and $\alpha_0$.
Accordingly the spectrum shows two separated peaks symmetrically around $\omega=0$.
For $\alpha_0 > {2}^{-1/2}$ the potential splits and shows three extrema: a maximum at $\alpha = 0$ with $V(0) = 0$ and two minima at $\alpha = \pm \sqrt{\frac{2\alpha_0^2 - 1}{2}}$.
For ${2}^{-1/2} < \alpha_0 < 1$ (\cref{fig5} (b)) the initial condition is above the maximum, such that $\alpha$ still oscillates symmetrically between $-\alpha_0$ to $+\alpha_0$, however the velocity around $\alpha=0$ is reduced, such the oscillation becomes strongly anharmonic.
The spectrum still consists of two main peaks symmetrically around zero and additional higher harmonics of these peaks which are, however, outside the plotted range. 
The transition point is reached at $\alpha_0 = \alpha_\te{t} = 1$, where the unstable fixed point $\alpha = 0$ (\cref{fig5} (c)) would be reached after infinite time.
The corresponding spectrum would be continuous, because no oscillation occurs.
Due to unavoidable numerical errors, however, this limiting case will in general not be exactly reached in numerical simulations.
Instead, after some rather long time either the barrier will be overcome, leading to a symmetric oscillation, or the motion turns around before reaching the unstable fixpoint, leading to an oscillation only in the range of positive alpha.
In both cases the spectrum consists of a series of equidistant peaks with small peak distance corresponding to the long oscillation period.
An example is shown on the right hand side of \cref{fig5} (c) where the simulation has been started nominally with $\alpha_0 = 1$.
The presence of a peak at $\omega=0$ indicates that this spectrum actually corresponds to an asymmetric oscillation, i.e., the case slightly above the threshold.
If $\alpha_0$ is further increased the dynamics take place only in the right valley and $\alpha$ stays positive for all times (\cref{fig5} (d)).
The offset gives rise to a peak at $\omega=0$, in addition to the two side peaks from the oscillation. The curvature at the minima $V''(\alpha = \sqrt{\frac{2\alpha_0^2 - 1}{2}}) = 2(2\alpha_0^2 - 1)$ increases with $\alpha_0$ and approaches the squared Rabi frequency $\Omega_\te{R}^2 = 4\alpha_0^2$ for high $\alpha_0$.
This explains the transition to the Rabi-splitting in \cref{fig3}, since the curvature equals the squared frequency of $\alpha$ in harmonic approximation.

As discussed above, the analytical model gives us the possibility to identify the transition point $\alpha_\te{t}$ exactly by calculating $V(0) = V(\alpha_\te{t})$, which is at $\alpha_\te{t}=1$.
Going back to the electric field we find that the transition takes place when
\begin{align}
 \alpha_\text{t} = \frac{\tilde{M}\tilde{E}_0}{\sqrt{\tilde{M}^2 \lambda}} = \frac{\tilde{E}_0}{\sqrt{\lambda}} = 1 \qquad
 \Leftrightarrow \tilde{E}_0 = \sqrt{\lambda} \sim \sqrt{N_\te{QD}/A} \,. \label{eq:transition}
\end{align}
Interestingly, the transition point depends only on system parameters like the quantum dot density $N_\te{QD}/A$, the cavity frequency $\omega_0$ or the refractive index at the position of the QDs $n(z_0)$, but does not depend on the dipole matrix element.

\section{Influence of a QD distribution with non-vanishing width}\label{sec:distribution}
\begin{figure}[t]
\centering
\includegraphics[width=\columnwidth]{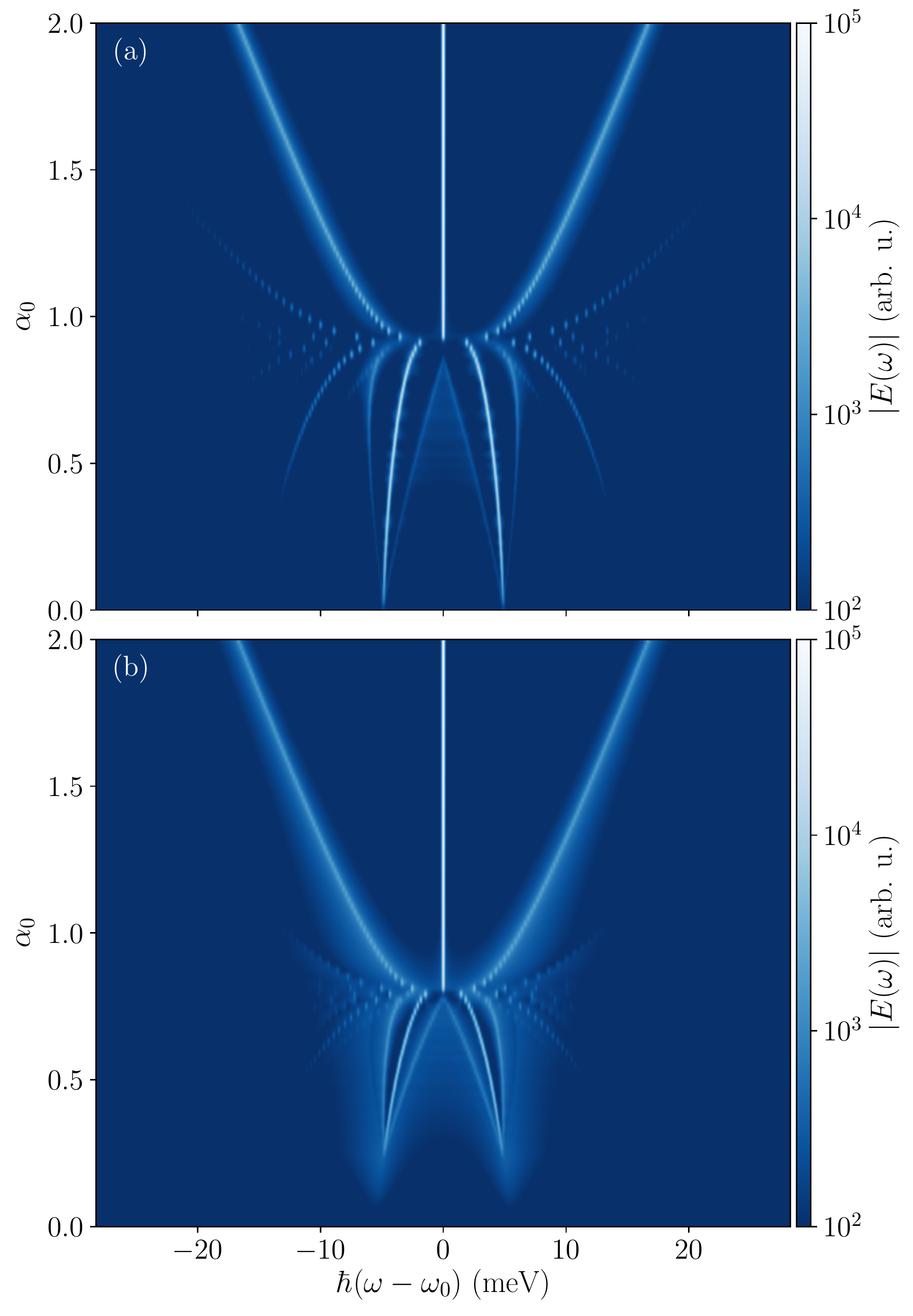}
\caption{Spectra $|E(\omega)|$ as functions of the initial field strength $\alpha_0$ for two different ensemble widths of (a) $\Delta_{\omega} = 4\,$meV and (b) $\Delta_{\omega} = 7\,$meV.}
\label{fig6}
\end{figure}

The assumption of a QD ensemble where all QDs have the same transition energy is strongly idealized.
In a real sample there will always be a certain spread of transition frequencies.
Therefore, in this section we will investigate how the spectra change when the cavity mode interacts with a QD ensemble with a Gaussian distribution of transition frequencies with FWHM $\Delta_{\omega}$ according to
\begin{align}
 \rho_\te{QD}(\omega) = \frac{2\sqrt{\ln(2)}}{\sqrt{\pi}\Delta_\omega} e^{-\frac{4\ln(2)(\omega - \omega_0)^2}{\Delta_{\omega}^2}}.
\end{align}
The influence of such kind of inhomogeneous broadening on the polariton spectra has been investigated in Refs.\cite{houdre1996vac,grochol2008mic} and an increasing broadening of the polariton lines with increasing width of the QD distribution has been found.
The question arises, whether also for increasing driving the spectra will just broaden and how the transition between the two regimes of low and high driving is affected by this broadened QD distribution.

Considering our analytical model (cf. \cref{eq:anal_eom}) and the nondimensionalization, the equations of motion change to
\begin{subequations}
 \begin{align}
 \dv{\tau} \alpha (\tau) &= \int \dd{\omega_x} \rho_\text{QD}(\omega_x) \gamma(\omega_x,\tau) \label{eq:a_alpha_e} \\
 \dv{\tau} \beta (\omega_x, \tau) &= -2 \left(\alpha^*(\tau) \gamma (\omega_x, \tau) + \alpha (\tau) \gamma^*(\omega_x, \tau)  \right) \label{eq:a_beta_e} \\
 \dv{\tau} \gamma(\omega_x, \tau) &= - i\frac{(\omega_x-\omega_0)}{\sqrt{\tilde{M}^2 \lambda}} \gamma(\omega_x, \tau) + \alpha(\tau) \beta(\omega_x, \tau)  \label{eq:a_gamma_e}.
\end{align}
\end{subequations}
Note that now $\alpha$ and  $\gamma$ become complex quantities.

From these equations we calculate the spectra of the electric field $|E(\omega)|$ depending on the initial value $\alpha_0$.
The results are shown in \cref{fig6} for two different values of the width of the QD distribution.
For a spectral width of $\Delta_{\omega} = 4\,$meV (\cref{fig6} (a)), similarly to \cref{fig3}  we find a sharp transition at $\alpha_0=\alpha_\te{t} \approx 1$ between a polaritonic and a Rabi oscillation regime.
This shows that the transition found in the limiting case of a $\delta$-like ensemble is also present for finite widths.
However, in contrast to a simple broadening of the spectral lines we observe the appearance of additional narrow spectral lines below the threshold.
Also for a larger value of $\Delta_{\omega} = 7$~meV in \cref{fig6} (b), the threshold $\alpha_\te{t}$ is still present, but shifted to a lower value $\alpha_\te{t} < 1$.
Below the threshold we find again additional lines; however, they appear only above another threshold $\alpha_\te{t2}$.
In the following we will try to understand this quite surprising dependence of the spectra on the QD distribution.

\begin{figure}[t]
\centering
\includegraphics[width=\columnwidth]{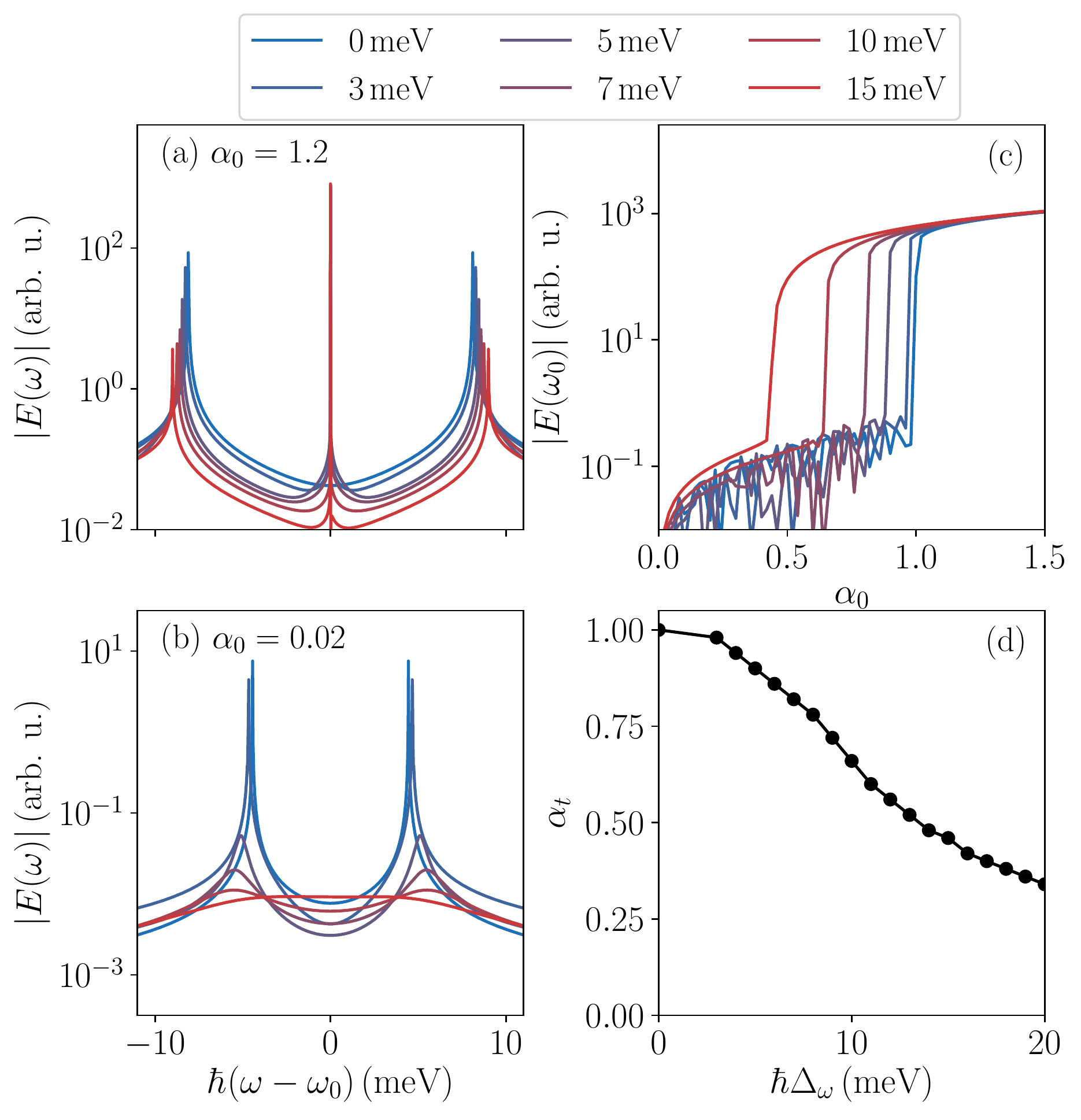}
\caption{Spectra $|E(\omega|$ for initial values (a) $\alpha_0 = 1.2$ and (b) $\alpha_0 = 0.02$  and (c) spectrum at the resonance frequency  $|E(\omega_0)|$ as a function of the initial value $\alpha_\te{t}$ for different ensemble widths as indicated above the plots.
(d) Transition point $\alpha_\te{t}$ as a function of the ensemble width $\Delta_{\omega}$.}
\label{fig7}
\end{figure}

Let us start with the polaritonic regime, i.e., the case of a weak initial field $\alpha_0 \ll 1$.
In this case, again the occupation can be neglected because it is quadratic in the field. Thus, $\beta \approx \beta_0 = -1$. The remaining equations (\ref{eq:a_alpha_e}) and  (\ref{eq:a_gamma_e}) are then linear equations which can be solved by a Fourier-Laplace transform.
The formal result is rather lengthy and not very instructive, therefore we do not present the formula.
However, one can deduce that with increasing broadening the splitting between the  polariton peaks slightly increases and the peaks broaden.
In addition, there is a broad maximum around $\omega = \omega_0$ which builds up with increasing broadening.
The spectra of the electric field at $\alpha_0=0.02$ for different values of the broadening are shown in \cref{fig7} (b).
Here, two clear peaks appear for small ensemble widths $\Delta_{\omega}$, but become broader and eventually vanish for larger $\Delta_{\omega}$.
For $\Delta_{\omega}=15$~meV only a single broad peak is visible.
These spectra are in good agreement with spectra found for microcavity polaritons in disordered exciton lattices obtained on the basis of a model with a quantized light field \cite{grochol2008mic}, demonstrating that indeed in the present case quantum features of the light are of minor importance.

Above the threshold $\alpha_\te{t}$ again the Rabi oscillation regime is reached.
As shown in \cref{fig7} (a) for the case of $\alpha_0=1.2$ , we obtain the three peak structure with a central peak at the cavity frequency and two side peaks split by the Rabi frequency.
Like in the polariton case, the splitting between the outer peaks slightly increases with increasing broadening.
This can be traced back to the fact that in a broadened ensemble there are QDs which are detuned from resonance and therefore exhibit a larger Rabi frequency than those at resonance.

The broadening of the polariton-like spectrum leads to the question, whether still a transition can be observed, even for ensembles with larger $\Delta_{\omega}$.
To answer this question, we show in \cref{fig7} (c) the spectrum at the cavity frequency $|E(\omega_0)|$ as function of the initial value $\alpha_0$.
For the $\delta$-like ensemble ($\Delta_{\omega}=0$) we find, as already discussed, a sharp increase of this amplitude at $\alpha_0=1$, which is also found for small ensemble widths.
Interestingly, also in the case of broad ensembles a sharp rise of the amplitude is found, indicating that still a transition between two regimes takes place.
We further observe that the transition $\alpha_\te{t}$ decreases to smaller values of $\alpha_0$ with increasing $\Delta_{\omega}$.
This is also quantified in \cref{fig7} (d), where we plot $\alpha_\te{t}$ as function of the ensemble width $\Delta_{\omega}$. 

As we have seen in the discussion of \cref{fig5} the transition is associated with the fact that $\alpha$ (in other words, the envelope of the electric field) does not reach zero anymore but oscillates only in the region of positive values.
This leads to the appearance of the central peak.
To demonstrate that this occurs necessarily also in the ensemble, we can introduce the total inversion $B(\tau)$ defined as
\begin{align}
 B(\tau) = \int \dd{\omega_x}  \rho_\text{QD}(\omega_x) \beta (\omega_x, \tau) .
\end{align}
With this definition we can combine \cref{eq:a_alpha_e,eq:a_beta_e} to
\begin{align}
 \dv{\tau} \left[2|\alpha (\tau)|^2 + B(\tau) \right] = 0.
\end{align}
Together with the initial conditions $\alpha(0)=\alpha_0$ and $B(0)=-1$ this leads to the energy conservation law
\begin{align}
2|\alpha (\tau)|^2 + B(\tau)  = 2\alpha_0^2 -1 \label{eq:alpha_B}
\end{align}
which is the direct generalization of \cref{eq:beta_int} to a broadened ensemble.
Since the inversion is limited to the range $-1 \le B(\tau) \le 1$, we obtain
\begin{align}
|\alpha(\tau)|^2 \ge \alpha_0^2 -1
\end{align}
This is a proof that for $\alpha_0 > 1$ the field envelope cannot vanish anymore and thus there is necessarily a peak at the cavity resonance $\omega_0$ in $|E(\omega)|$.
In a broadened ensemble $B=1$ is not reached, because this would mean that all QDs are completely inverted, no matter how far they are detuned from resonance.
This explains our findings of \cref{fig7} (c) and (d).

\begin{figure}[t]
\centering
\includegraphics[width=\columnwidth]{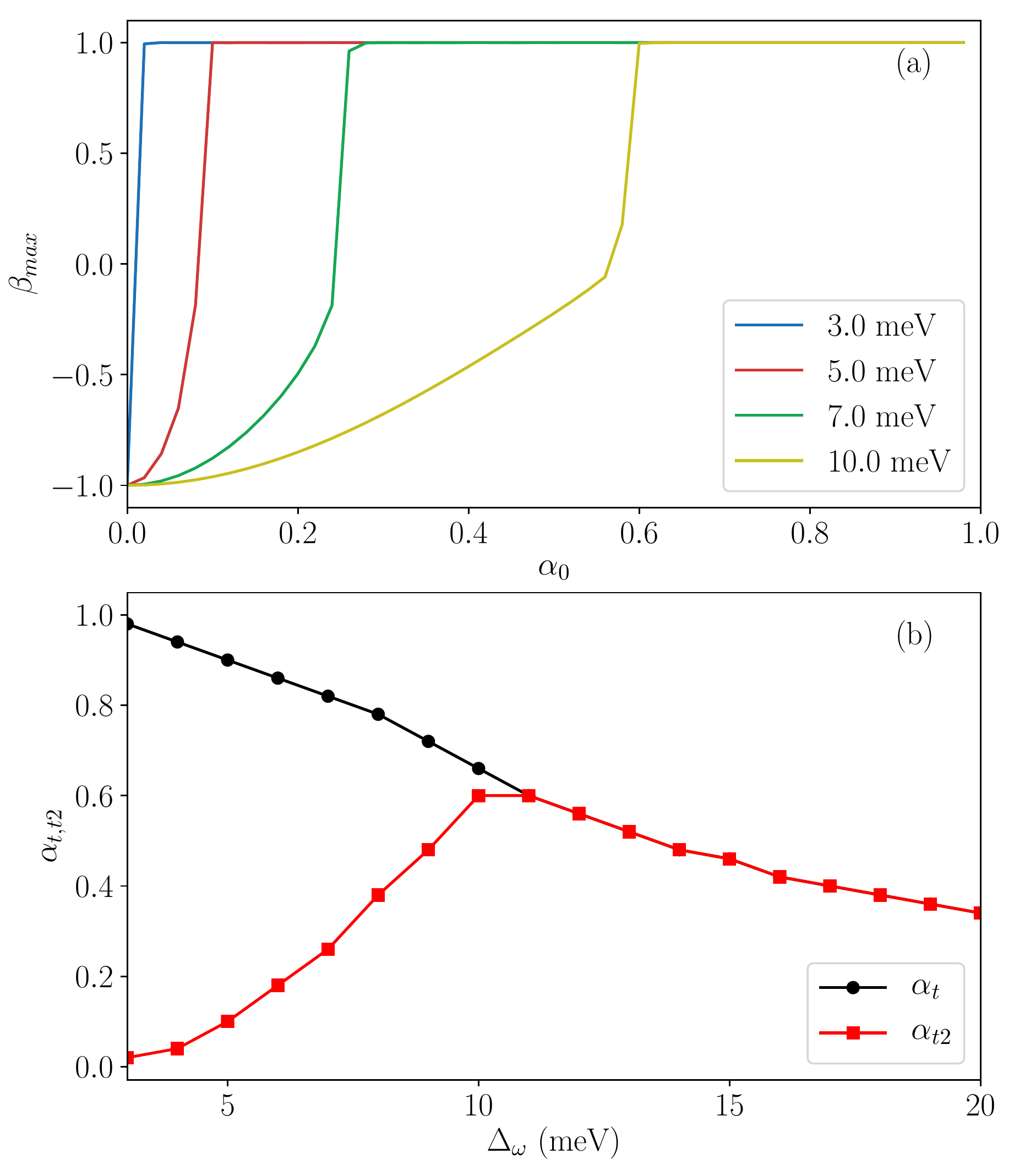}
\caption{(a) Maximum value of the inversion as a function of $\alpha_0$ for ensembles of different widths $\Delta_\omega$.
(b) Lower transition $\alpha_\te{t2}$ (red squares) where the inversion at the polariton peaks reaches unity and upper transition $\alpha_\te{t}$ (black circles) where the inversion at the cavity frequency reaches unity as functions of the ensemble width $\Delta_\omega$.}
\label{fig8}
\end{figure}

Finally we want to understand the origin of the additional lines in the spectra of \cref{fig6} starting from the polariton lines.
They are symmetric with respect to the polariton lines, which already indicates that they will originate from an additional modulation of the field amplitude.
Indeed it turns out that they are again related to Rabi oscillations, in contrast to the case above the threshold $\alpha_\te{t}$, however, these are not Rabi oscillations of the QDs in resonance with the cavity mode but Rabi oscillations of the QDs in resonance with the polariton frequencies.
\Cref{fig8} (a) shows the maximum of the inversion $\beta_\te{max}$ that is reached in the ensemble as a function of the initial field amplitude $\alpha_0$ for different values of the ensemble width.
The maximum inversion increases with increasing $\alpha_0$ until it reaches unity. It turns out that this maximum is indeed reached at the polariton frequencies, since there the driving of the QDs is strongest.
As soon as the maximum inversion reaches unity, Rabi oscillations of the QDs in resonance with the polaritons set in leading to a modulation of the electric field and thus the observed side peaks of the polaritons.
The splitting between these side peaks increases with $\alpha_0$, as in the case of the Rabi oscillation above $\alpha_\te{t}$. The onset of these polariton Rabi oscillations, i.e., the point where the inversion reaches unity, occurs at a second, lower threshold $\alpha_\te{t2}$ which increases with increasing ensemble width, as is shown in \cref{fig8} (b) together with the upper threshold $\alpha_\te{t}$ already discussed in \cref{fig7} (d).
This increase reflects the fact that the number of QDs at the polariton frequency increases with increasing ensemble width, such that a larger initial field is required to invert all these QDs.
At a width of about 10~meV the two thresholds merge.
Here the broadening of the polariton lines becomes so strong (see \cref{fig7} (b)) that an inversion of unity is first reached at the cavity frequency, which defines the threshold $\alpha_\te{t}$.

\section{Conclusion}
In this work we have calculated the dynamics of a combined QD-cavity system driven by an external laser pulse in a semiclassical model.
For the numerical calculations we employed an FDTD method with incorporated two-level systems. First we studied a system with identical QDs having transition energies in resonance with the cavity mode.
Depending on the excitation power we found a fundamentally different behavior of the spectrum of the QD-cavity system when varying the driving strength: For low pulse amplitudes we found exciton-polariton-like states with two amplitude-independent peaks in the spectrum.
High pulse areas resulted in typical Rabi oscillations, where the cavity mode has the most dominant part in the spectra and the field induced by the QDs is small.
Accordingly, the spectrum shows three peaks, where the side peaks are given by the Rabi splitting.
Interestingly, in between these two regimes there is a sharp transition.
We identify the sharp transition in the numerical model by the point, where the envelope of the electric field does not reach zero anymore.

We furthermore have derived an analytical model resulting in a set of nonlinear coupled equations, which describe the dynamics found in almost perfect agreement with the FDTD simulation.
Our analytical model allowed us to interpret the different regimes.
In particular, we were able to explain the transition by showing that the field dynamics can be mapped to Newton-like dynamics in a fourth order potential.
The shape of the potential depends on the initial condition, changing from a single well potential for weak driving to a double well potential for strong driving.
The transition occurs when the driving reaches a value such that the oscillations remain in one of the two minima and thus, as found in the FDTD simulations, when the field amplitude does not reach zero anymore. Close to the transition the dynamics are strongly anharmonic leading to higher harmonics in the spectrum. 

We finally extended the analytical model to account for finite ensemble widths and showed that at low driving the polariton lines broaden and slightly shift.
However, the sharp transition to the Rabi oscillation regime exists independent of the ensemble width.
Interestingly, instead of seeing just a broadening, in the polariton regime we observe new lines above a second, lower threshold, which could be traced back to Rabi oscillations of the QDs in resonance with the polariton lines.
Our paper thus provides an intuitive model of the coupled cavity-QD system with and without broadening as a prototypical example of a nonlinear system.

\appendix

\section{Reduction to the circularly polarized light field} \label{app:circular}

According to \cref{eq:E_circ} a $\sigma_{+}$ ($\sigma_{-}$) circularly polarized light field with central frequency $\omega_0$ is given by
\begin{align}
\vect{E}(z,t) = \frac{1}{\sqrt{2}} \left[ E\left(z,t\right) \vect{e}_x \pm E\left(z,t-\frac{\pi}{2\omega_0}\right) \vect{e}_y \right]. 
\end{align}
Separating the field in its positive and negative frequency components according to
\begin{align}
E\left(z,t\right) = \frac{1}{2} \left[ \tilde{E}(z,t) e^{-i \omega_0 t} + \tilde{E}^*(z,t) e^{i \omega_0 t}\right]
\end{align}
the field reads
\begin{align}
\vect{E}(z,t) &= \frac{1}{2\sqrt{2}} \biggl\{ \left[ \tilde{E}\left(z,t\right) e^{-i \omega_0 t}+ \tilde{E}^*\left(z,t\right) e^{i \omega_0 t} \right] \vect{e}_x  \nonumber \\  
& \qquad \pm \left[ \tilde{E}\left(z,t-\frac{\pi}{2\omega_0}\right) e^{-i \omega_0 t +i\frac{\pi}{2}} \right. \nonumber \\ & \left. \qquad \quad + \tilde{E}^*\left(z,t-\frac{\pi}{2\omega_0}\right) e^{i \omega_0 t -i\frac{\pi}{2}} \right] \vect{e}_y \biggr\} .
\end{align}
Assuming that the envelope is slowly varying on the oscillation period of the field, i.e., 
\begin{align}
\tilde{E}\left(z,t\right) \approx \tilde{E}\left(z,t - \frac{\pi}{2\omega_0}\right) ,
\end{align}
the real electric field can be written as
\begin{align}
\vect{E}(z,t) &= \frac{1}{2\sqrt{2}} \left[ \left( \vect{e}_x \pm i\vect{e}_y \right)\tilde{E}\left(z,t\right) e^{-i \omega_0 t} \right. \nonumber \\ 
& \left. \qquad \quad +  \left( \vect{e}_x \mp i\vect{e}_y \right)\tilde{E}^*\left(z,t\right) e^{i \omega_0 t} \right] \nonumber \\ 
&= \frac{1}{2} \left[  \vect{e}_\pm \tilde{E}\left(z,t\right) e^{-i \omega_0 t}+  \vect{e}_\mp \tilde{E}^*\left(z,t\right) e^{i \omega_0 t} \right] .
\end{align}
This shows that indeed the positive frequency component is proportional to the polarization vector $\vect{e}_\pm$.

The macroscopic polarization of the $\sigma_{+}$ ($\sigma_{-}$) circularly polarized exciton in a QD reads (for simplicity here we omit the superscript $(n)$ labeling the QD)
\begin{align}
\vect{P}(t) &=  M \left[ \vect{e}_\pm p + \vect{e}_\mp p^* \right] \nonumber \\
&= \frac{1}{\sqrt{2}} M \left[ \vect{e}_x \left( p + p^* \right) \pm i \vect{e}_y \left( p - p^* \right) \right]\, .
\end{align}
Using
\begin{align}
p(t) = \tilde{p}(t) e^{-i\omega_x t}
\end{align}
and assuming again a slowly varying envelope 
\begin{align}
\tilde{p}(t) \approx \tilde{p}\left(t-\frac{\pi}{2\omega_x}\right)
\end{align}
we have
\begin{align}
ip(t) &= \tilde{p}(t) e^{-i\omega_x t+i\frac{\pi}{2}} \approx p\left( t-\frac{\pi}{2\omega_x} \right) \\
-ip^*(t) &= \tilde{p}^*(t) e^{i\omega_x t-i\frac{\pi}{2}} \approx p^*\left( t-\frac{\pi}{2\omega_x} \right) .
\end{align}
This leads to
\begin{align}
\vect{P}(t) &= \frac{1}{\sqrt{2}} M \biggl\{ \vect{e}_x \biggl[ p(t) + p^*(t) \biggr] \nonumber \\
& \pm \vect{e}_y \left[ p\left( t-\frac{\pi}{2\omega_x} \right) + p^*\left( t-\frac{\pi}{2\omega_x} \right) \right] \biggr\}
\end{align}
showing that with
\begin{align}
P(t) = M \left[ p(t) + p^*(t) \right]
\end{align}
we obtain
\begin{align}
\vect{P}(t) &= \frac{1}{\sqrt{2}}  \left[ P(t) \vect{e}_x \pm P\left( t-\frac{\pi}{2\omega_x} \right) \vect{e}_y \right],
\end{align}
which has the same structure as  \cref{eq:E_circ} for the electric field.
Therefore, $P(t)$ is the source for the field $E(t)$.


%

\end{document}